\documentclass[apj]{emulateapj}

\newcommand{\niv}{N\,{\sc iv]}}
\newcommand{\niii}{N\,{\sc iii]}}
\newcommand{\lya}{Ly$\alpha$}
\newcommand{\nv}{N\,{\sc v}}
\newcommand{\oi}{O\,{\sc i}}
\newcommand{\siii}{Si\,{\sc ii}}
\newcommand{\siiv}{Si\,{\sc iv}}
\newcommand{\civ}{C\,{\sc iv}}
\newcommand{\heii}{He\,{\sc ii}}
\newcommand{\oiii}{O\,{\sc iii]}}
\newcommand{\aliii}{Al\,{\sc iii}}
\newcommand{\siiii}{Si\,{\sc iii]}}
\newcommand{\ciii}{C\,{\sc iii]}}
\newcommand{\mgii}{Mg\,{\sc ii}}
\newcommand{\feii}{Fe\,{\sc ii}}
\newcommand{\lsim}{\stackrel{\scriptscriptstyle <}{\scriptstyle {}_\sim}}

\begin{document}

\title{A SAMPLE OF QUASARS WITH STRONG NITROGEN EMISSION LINES 
FROM THE SLOAN DIGITAL SKY SURVEY}

\author{Linhua Jiang\altaffilmark{1}, Xiaohui Fan\altaffilmark{1},
and M. Vestergaard\altaffilmark{2}}
\altaffiltext{1}{Steward Observatory, University of Arizona, Tucson, AZ 85721}
\altaffiltext{2}{Department of Physics and Astronomy, Tufts University,
Medford, MA 02155}

\begin{abstract}

We report on 293 quasars with strong \niv\ $\lambda$1486 or \niii\ 
$\lambda$1750 emission lines (rest-frame equivalent width $>3$ \AA) at 
$1.7<z<4.0$ selected from the Sloan Digital Sky Survey (SDSS) Fifth Data
Release. These nitrogen-rich (N-rich) objects comprise
$\sim1.1$\% of the SDSS quasars. The comparison between the N-rich quasars and 
other quasars shows that the two quasar subsets share many common properties.
We also confirm previous results that N-rich quasars have much stronger \lya\ 
and \nv\ $\lambda$1240 emission lines. Strong nitrogen emission in all 
ionization states indicates high overall nitrogen abundances in these objects. 
We find evidence that the nitrogen abundance is closely related to quasar 
radio properties. The radio-loud fraction in the \niii-rich quasars is 26\% 
and in the \niv-rich quasars is 69\%, significantly higher than $\sim8$\% 
measured in other quasars with similar redshift and luminosity. Therefore, the 
high nitrogen abundance in N-rich quasars could be an indicator of a special 
quasar evolution stage, in which the radio activity is also strong.

\end{abstract}


\section{INTRODUCTION}

Quasars with prominent nitrogen emission lines at 1486 \AA\ or 1750 \AA\ are 
rare. A well known example is Q0353-383, a luminous quasar showing strong 
\niv\ $\lambda$1486, \niii\ $\lambda$1750, and \nv\ $\lambda$1240 emission 
lines \citep{osm80,bal03}. \citet{bal03} claimed that the unusual nitrogen 
emission in Q0353-383 is likely due to high metallicity in the broad line 
region of the quasar; the metallicity measured from line strength ratios 
involving \nv, \niv, and \niii\ is $\sim$15 times the solar abundance.

The Sloan Digital Sky Survey \citep[SDSS;][]{yor00} provides statistically
significant samples to study these nitrogen-rich (N-rich) quasars. From a 
sample of $\sim5600$ objects in the SDSS First Data Release Quasar Catalog 
\citep{sch03}, \citet{ben04} reported on 20 N-rich quasars in the redshift 
range $1.6<z<4.1$. The fraction of N-rich quasars in their sample is about 
0.4\%. By comparing with other quasars, they found that the N-rich quasars 
tend to have stronger \lya\ emission lines, and stronger but narrower \civ\ 
$\lambda$1549 and \ciii\ $\lambda$1909 emission lines. 
The two quasars with the strongest \niv\ and \niii\ lines in the 
\citet{ben04} sample have been studied in detail by \citet{dha07}. Recently 
\citet{gli07} discovered two quasars with strong \niv\ lines from a sample of 
23 faint quasars at $3.7<z<5.1$. They suspected that the high detection rate 
of N-rich quasars is due to the low luminosity and high redshift of the sample.

The number of known N-rich quasars is still small, and most of their 
properties remain unclear, such as the UV/optical continuum slopes, emission 
line strengths and widths, and central black hole masses, etc. In this paper, 
we present 293 N-rich quasars at $z>1.7$ selected from the SDSS Fifth Data 
Release (DR5) Quasar Catalog \citep{sch07}. In $\S\,2$, we introduce our sample
selection. We measure various properties of these quasars and compare them
with those of other SDSS quasars in $\S\,3$. A short summary is given in 
$\S\,4$. Throughout the paper we use a 
$\Lambda$-dominated flat cosmology with H$_0=70$ km s$^{-1}$ Mpc$^{-1}$, 
$\Omega_{m}=0.3$, and $\Omega_{\Lambda}=0.7$ \citep{spe07}. We use EW and FWHM 
to denote {\it rest-frame} equivalent width and full width at half maximum, 
respectively.

\section{SAMPLE SELECTION}

The SDSS is an imaging and spectroscopic survey using a dedicated wide-field
2.5m telescope \citep{gun06}. The imaging is carried out in five broad bands,
$ugriz$, spanning the range from 3000 to 10,000 \AA\ \citep{fuk96};
spectroscopy is performed using a pair of double spectrographs with coverage
from 3800 to 9200 \AA, and a resolution $\lambda/\Delta \lambda$ of roughly
2000. The SDSS quasar survey spectroscopically targets quasars with $i<19.1$
at low redshift ($z\le3$) and $i<20.2$ at high redshift ($z\ge3$).
Quasar candidates
are mostly selected in $ugri$ and $griz$ color space \citep{ric02}. An SDSS
object is also considered to be a primary quasar candidate if it is located
within $2\arcsec$ of a radio source of the Faint Images of the Radio Sky at
Twenty-cm survey \citep[FIRST;][]{bec95}. The sample we use is from the SDSS 
DR5 Quasar Catalog. The catalog consists of 77,429 quasars with luminosities 
larger than $M_{i}=-22$ \citep{sch07}.

To find quasars with prominent \niv\ or \niii\ emission lines, we searched
25,941 quasar spectra with $i<20.1$ in the redshift range $1.7<z<4.0$. The 
search was done by visual inspection of each spectrum. This step results in a 
sample of 380 N-rich quasar candidates. We then measured the rest-frame 
EW and FWHM for each detected 
\niv\ or \niii\ line as follows. We estimate the local continuum by fitting a 
power-law to the spectra at both sides of the line. After subtraction of this 
continuum, a Gaussian profile is fitted to the line, and the EW and FWHM are 
computed from the best-fitted component. Our final selection criterion of 
N-rich quasars is $\rm EW>3$ \AA. This cut is chosen for the following 
reasons: (1) In most quasar spectra, the significance of the detection of a
EW = 3 \AA\ nitrogen line is greater than $5\sigma$; (2) The 3 \AA\ cut is 
significantly larger than typical EW values for the two nitrogen lines
(see $\S\,3.2$); (3) \citet{ben04} used a similar
selection cut. A total of 293 quasars in our sample meet this criterion, 
including 43 quasars with strong \niv\ lines and 279 quasars with strong 
\niii\ lines (29 quasars have both lines). Table 1 presents the catalog of
the N-rich quasars. Column (1) gives the name of each quasar, and column (2)
lists the redshift. Column (3) shows the SDSS $i$-band magnitude corrected for 
Galactic extinction. All other quantities will be explained in $\S$ 3. 
The full Table 1 appears in the electronic edition of the {\it Astrophysical
Journal}.

Figure 1 shows four good examples of the selected quasars. Figure 2 shows the
rest-frame EW and FWHM distributions of \niii. Figure 3 compares the redshift 
distributions of the N-rich quasars and SDSS DR5 quasars at $z>1.7$. The 
distributions are consistent at most redshift ranges except for the ranges of 
$2.5<z<2.9$ and $z>3.5$. At $2.5<z<2.9$, the higher fraction of the N-rich 
quasars is caused by the higher SDSS quasar selection efficiency due to 
stronger \lya\ emission in these quasars ($\S\,3.2$). The lower fraction of 
the N-rich quasars at $z>3.5$ is likely due to the lower signal-to-noise 
ratios (SNRs) of the spectra at $\lambda\ge 8000$ \AA.

\section{RESULTS}

We detect 293 quasars with prominent \niv\ or \niii\ emission lines from 
25,941 SDSS quasars at $1.7<z<4.0$. This is the largest N-rich quasar sample 
to date. The fraction of N-rich quasars is about 1.1\%, significantly higher 
than 0.4\% (20 out of 5600) found by \citet{ben04}, who used a similar EW 
selection cut in a similar redshift range. This is likely because detecting
weak lines in spectra of moderate SNRs is difficult for automated codes.
Based on repeated selections of N-rich quasars by visual inspection from a 
subset of 2000 quasars, we estimate 
that the completeness at $\rm EW>3$ \AA\ is greater than 80\%. In this section 
we compare various properties of the N-rich quasars with those of `normal' 
SDSS quasars.

\subsection{Continuum and Emission Line Properties of N-rich Quasars}

To analyze the emission line properties of each N-rich quasar, we fit a 
power-law ($f_{\nu}\propto\nu^{\alpha}$) to regions with very little 
contribution from line emission. The quasar spectrum was first scaled to its 
$r$-band magnitude and thereby placed on an absolute flux scale. After fitting
and subtraction of the power-law continuum, we fit the following individual 
emission lines. (1) \siiv\ $\lambda$1396 if $i<19.0$ and $z>1.9$. We use a 
single Gaussian profile to fit this line. (2) \civ\ if $i<19.5$. We use a 
single Gaussian and double Gaussians to fit this line respectively, and we 
choose the best fit. (3) \ciii\ if $i<19.5$ and $z<3.5$. The weak \aliii\
$\lambda1857$ line is often detected in the blue wing of the \ciii\ line. We 
use two Gaussian profiles to fit these two lines simultaneously if \aliii\ is 
apparent, otherwise \ciii\ is fitted using a single Gaussian component. 
We also note that there is usually a contamination from the weak \siiii\ 
$\lambda1892$ line. We do not remove this contamination because the 
decomposition of the \ciii\ and \siiii\ lines is difficult and the strength of
\siiii\ is minor compared to \ciii\. 
All models are visually inspected, and the emission lines 
severely affected by absorption lines are discarded from our analysis. The 
EW and FWHM are measured from the best fits. The results are shown
in Table 1. Column 2 of Table 2 lists the median values 
of EW and FWHM for each line. The median value of the power-law slopes is 
also given in Table 2. The slope is slightly redder than the slope in the 
\citet{van01} composite spectrum ($alpha_\nu =-0.44$). In most cases the 
spectral quality is not good enough to allow more sophisticated fittings as 
\citet{bal03} and \citet{dha07} did, or to allow reliable modeling of
weak emission lines such as \heii\ $\lambda$1640 and \oiii\ $\lambda1663$.
The \lya, \nv, and \siii\ $\lambda1262$ lines are usually heavily blended, so 
modeling their profiles requires high quality spectra. We discuss their 
average properties in $\S\,3.2$.

In the above analysis we do not consider the UV \feii\ emission that 
contaminates most of the UV spectral region. The most prominent \feii\ 
emission line complex is at 2200\AA{} $\lsim \lambda \lsim$ 3000 \AA\ 
\citep{ves01}. At
$\lambda<2200$ \AA, the \feii\ emission is weak, and does not significantly 
affect the measurements of strong emission lines. At $z>2.1$, the 
strong \feii\ complex is moving out of the SDSS spectral coverage, making it 
difficult to measure the \feii\ emission. However, Fe abundance is of 
interest, and the \feii/\mgii\ $\lambda$2800 ratio is important for
understanding chemical abundances in quasars \citep[e.g.][]{ham99,die02}. 
We fit and measure the \feii\ emission for quasars with $i<19.0$ and $z<2.1$.
The modeling is done as an iteration over a power-law fit to the 
continuum emission and a model fit to the \feii\ emission using a scaled and 
broadened version of the \feii\ template of \citet{ves01}. The \feii\ template
is broadened by convolving the template with Gaussians with a range of sigmas.
In the first iteration a power-law is fitted to the continuum windows.
Upon subtraction of this primary continuum fit multiple
broadened copies of the \feii\ template are scaled to regions which
predominantly contain Fe emission. The broadened and scaled \feii\ template
that provides the best fit is selected and subtracted. Then another power-law
fit to the continuum emission is performed. The iteration of the continuum and
\feii\ fitting is repeated until convergence is obtained. See \citet{ves01} 
for the detailed process. All models are visually inspected, and poor fits
(due to low SNRs or strong absorptions) are discarded from our analysis.

After the subtraction of the \feii\ emission and power-law continuum we fit
the \mgii\ line using a Gaussian profile and measure its EW and FWHM from 
the best fit. We calculate the flux of the \feii\ complex by integrating
the best-fitting \feii\ template from 2200 to 3090 \AA\ \citep[e.g.][]{die02}.
The results are shown in Table 1, and their mean values are shown in Table 2.
The mean value of the \feii/\mgii\ ratios is 
3.6 with a scatter 1.3, consistent with previous measurements in normal
quasars \citep[e.g.][]{iwa02}.

\subsection{Comparison Between N-rich Quasars and `Normal' Quasars}

To find differences between N-rich quasars and `normal' quasars, we draw a
quasar sample three times larger than the sample of the N-rich quasars from
the SDSS DR5 quasar catalog by random selection. 
The sample has $i$-band magnitudes brighter than 20.1 and redshift 
distribution similar to that of the N-rich sample shown in 
Figure 3. We measure continuum and emission line properties for this sample 
using the methods outlined in $\S\,3.1$. The results are given in Column 3 of 
Table 2. The comparison in Table 2 shows that most of the properties in the 
two samples are very consistent, including the continuum slopes, the 
\feii/\mgii\ ratios, and the EWs and FWHM of most 
emission lines. The only apparent difference is the FWHM of \civ\ and \ciii; 
their values in the N-rich sample is about 25--30\% smaller than those in the 
SDSS DR5 sample. This was also noticed by \citet{ben04}.

We compare the luminosity distributions of the two quasar samples. We measure 
$m_{2500}$, the apparent AB magnitude at rest-frame 2500 \AA, based on
$m_{2500}=-2.5\times {\rm log}\,(f_{2500})-48.60$, where $f_{2500}$ is the 
monochromatic flux density (in units of $\rm erg\, sec^{-1}\, cm^{-2}\, 
Hz^{-1}$) at rest-frame 2500 \AA. The flux density $f_{2500}$ is derived
using the best-fitted continuum. We then determine the absolute magnitude 
$M_{2500}$ from $m_{2500}$. We find that the distributions of 
$M_{2500}$ for the two samples are similar, although the N-rich sample is 
about 0.5 magnitude brighter on average. This is caused by our selection; 
we may have missed some faint N-rich quasars where SNRs are low and nitrogen 
lines are difficult to identify. However, the difference of 0.5 magnitude is 
too small for the Baldwin effect \citep{bal77} to impact the comparison of the
EWs.

To compare the overall properties of continua and emission lines, we construct
composite spectra for the two samples. Each spectrum is placed in the rest 
frame and scaled so that its
continuum at 1450 \AA\ is equal to 1 (arbitrary units). The scaled spectra
are then averaged without any weights. This is to avoid bias introduced by
the Baldwin effect if we use weights involving SNRs. The two composite spectra 
are shown in Figure 4. They have similar shapes from 1000 to 3000 \AA. By 
definition the \niv\ and \niii\ lines in the N-rich spectrum are stronger.
In the SDSS DR5 spectrum, the EWs of \niii\ is $0.40\pm0.06$ \AA, consistent
with $0.44\pm0.03$ \AA\ in the \citet{van01} composite spectrum; the EWs of 
\niv\ is $-0.04\pm0.06$ \AA, consistent with no detection.
We measure the properties of \lya\ and \nv\ as follows. After the subtraction 
of the power-law continuum, we fit the \lya, \nv, and \siii\ lines 
simultaneously using four Gaussian profiles, with the first two profiles 
representing the broad and narrow components of \lya\ and the last two 
profiles representing \nv\ and \siii. For \lya, we only fit its red half side
and mirror the fit around the peak for the EW and FWHM measurements.
The EWs and FWHM for the lines are determined using the methods
described in $\S\,3.1$. The results are shown in Table 2. Errors in these 
measurements are negligible owing to the high SNRs of the composite spectra.
Compared to normal quasars, the N-rich quasars have $\sim40$\% stronger 
\lya\ and \nv\ emission, and $\sim100$\% stronger \oi\ $\lambda$1304 
emission. Their \civ\ and \ciii\ line widths are 20--25\% smaller, as we 
find in $\S\,3.1$.

The radio-loud fraction (RLF) of optically-selected quasars is a strong 
function of redshift and optical luminosity. We calculate the RLF for our
N-rich sample following \citet{jia07}. Radio loudness $R$ is defined by
$R=f_{6cm}/f_{2500}$, where $f_{6cm}$ is the observed flux density at 
rest-frame 6 cm derived from FIRST (if detected) assuming a power-law slope of 
$-0.5$. When quasars with $R>10$ are defined 
as radio-loud, FIRST is able to detect radio-loud quasars down to 
$i\approx 18.9$ \citep{jia07}. We find that the RLF in the \niii-rich quasars 
is 26\% and in the \niv-rich quasars is 69\%. The RLF in quasars with both 
\niv\ and \niii\
lines even reaches 80\%, significantly higher than the RLF $\sim8$\% found in
normal SDSS quasars with similar redshift and luminosity \citep{jia07}.
This suggests that the \niv\ and \niii\ emission in these quasars are related 
to the quasar radio properties.

Central black hole masses of quasars can be estimated using mass scaling 
relations based on broad emission line widths and continuum luminosities. The 
scaling relations are written as $\rm M_{BH}\propto FWHM^2\,L_{opt}^\beta$, 
where $\rm L_{opt}$ is optical luminosity and $\beta\approx0.5$. Strong 
emission lines such as \mgii\ and \civ\ have been widely used to determine 
black hole masses \citep[e.g.][]{mcl04,ves06,she07}. 
Based on the similar luminosities and FWHM of \mgii\ in Table 2,
the black hole masses in the two quasar subsets are consistent.

Photoionization models have shown that a series of emission-line ratios can
be used to estimate gas metallicity in the broad line region of quasars 
\citep{ham02,nag06}. In particular, the relative nitrogen abundance is a good 
metallicity indicator if it serves as a secondary element so that 
N/O$\propto$O/H. By definition, N-rich quasars have much stronger 
\niv\ or \niii\ emission. They also have stronger \lya\ and \nv\ emission.
Since \nv, \niv, and \niii\ have a range of critical densities and ionization
parameters, stronger \nv, \niv, and \niii\ emission shows higher nitrogen 
abundance, and higher metallicities if the relation N/O$\propto$O/H holds in 
these quasars. For example, a $\sim40$\% stronger \nv/\civ\ ratio in N-rich 
quasars in Table 2 indicates a $\sim40$\% higher metallicity according to 
\citet{ham02}. However, the strengths of most other emission lines in N-rich 
quasars are similar to those in normal quasars, and thus the metallicities 
measured from line ratios other than nitrogen \citep{nag06} would be 
consistent for the two samples, which is contrary to the results 
from nitrogen. Therefore, stronger \nv, \niv, and \niii\ emission in N-rich 
quasars may not indicate a higher overall metallicity; instead, it simply 
means a larger nitrogen abundance. The nitrogen enrichment in these quasars 
could significantly deviate from the usual N/O$\propto$O/H scaling. 

\section{SUMMARY}

We have obtained a sample of 293 SDSS quasars with strong \niv\ or \niii\ 
emission lines at $z>1.7$. The fraction of N-rich quasars is about 1.1\%. 
The comparison between these N-rich quasars and normal quasars shows that
the two quasar subsets have similar: (1) redshift and luminosity 
distributions; (2) UV continuum slopes; (3) line strengths and widths for most 
emission lines; and (4) central black hole masses. 
However, the N-rich quasars have:
(1) $\sim40$\% stronger \lya\ and \nv\ emission. Strong nitrogen emission in 
all ionization states indicates high overall nitrogen abundances. (2) 
$20-25$\% narrower widths of the \civ\ and \ciii\ lines; and (3) a much higher 
RLF. Almost all the quasars with both \niv\ and \niii\ lines are radio-loud.

The most intriguing difference between the two quasar subsets is the much 
higher RLF of N-rich quasars. We notice that Q0353-383 is also a radio-loud 
quasar with $R\approx130$ based on the optical observation of \citet{osm80} 
and the radio observation of \citet{con98}. Also the \niv\ and \niii\ lines 
are rarely seen in normal galaxies, but they seem more common in luminous 
radio galaxies with AGN signatures. For example, the two lines were detected 
at $\ge5\sigma$ significance in 67\% radio galaxies in the sample of 
\citet{ver01}. All these results indicate that the nitrogen abundance is 
tightly related to quasar radio properties: perhaps the high nitrogen 
abundance in N-rich quasars is an indicator of a special quasar evolution 
stage, in which the radio activity is also very strong. 
Detailed models, which are beyond the scope of this letter, have to explain 
all the distinguishing \mbox{features of
N-rich quasars found in this work.}

\acknowledgments

We acknowledge support from NSF grant AST-0307384, a Sloan Research Fellowship
and a Packard Fellowship for Science and Engineering (LJ, XF). We thank 
F. Hamann, C. Tremonti and D. Zaritsky for helpful discussions.

Funding for the SDSS and SDSS-II has been provided by the Alfred P. Sloan
Foundation, the Participating Institutions, the National Science Foundation,
the U.S. Department of Energy, the National Aeronautics and Space
Administration, the Japanese Monbukagakusho, the Max Planck Society, and the
Higher Education Funding Council for England. The SDSS is managed by the
Astrophysical Research Consortium for the Participating Institutions.
The SDSS Web Site is http://www.sdss.org/.

\clearpage
\begin{figure}
\epsscale{0.6}
\plotone{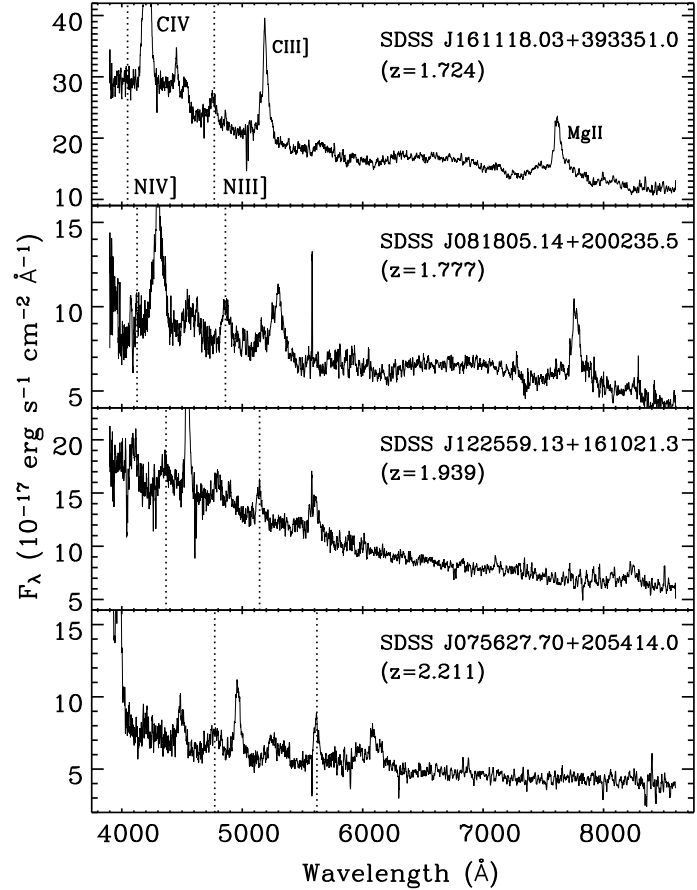}
\caption{Spectra of four N-rich quasars, smoothed by a boxcar of 5
   pixels. The dotted lines indicate the positions of \niv\ and \niii.
   The first two quasars show strong \niii\ lines and the last two quasars
   show both \niv\ and \niii\ lines.}
\end{figure}

\begin{figure}
\plotone{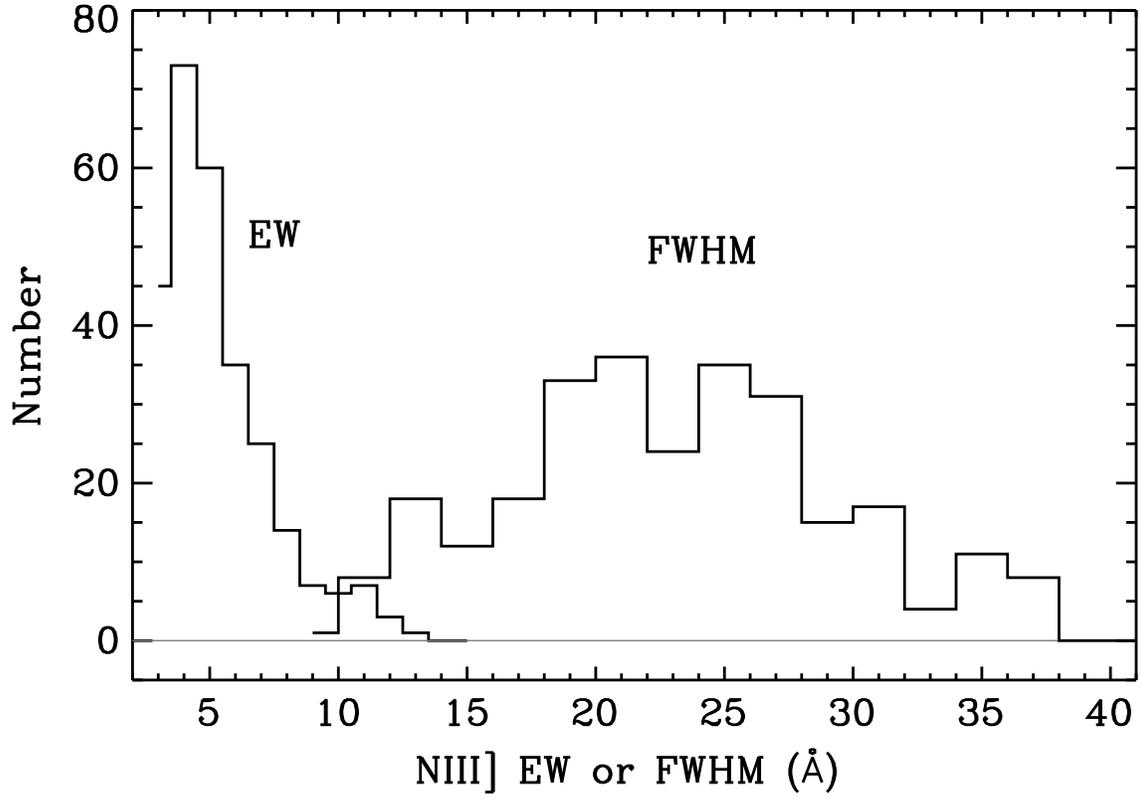}
\caption{Rest-frame EW and FWHM distributions of \niii.}
\end{figure}

\begin{figure}
\plotone{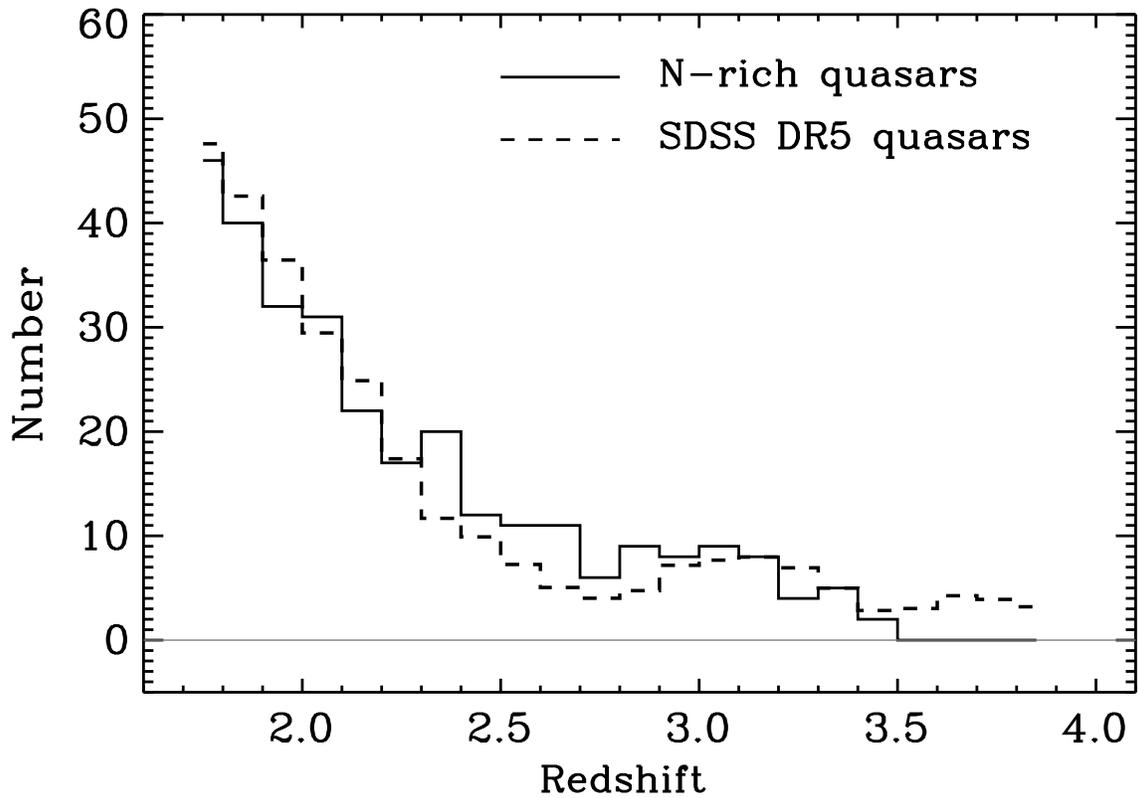}
\caption{Redshift distributions of the N-rich quasars and SDSS DR5 quasars.
   The histogram for the SDSS DR5 quasars has been scaled to match the
   number of the N-rich quasars.}
\end{figure}

\begin{figure}
\plotone{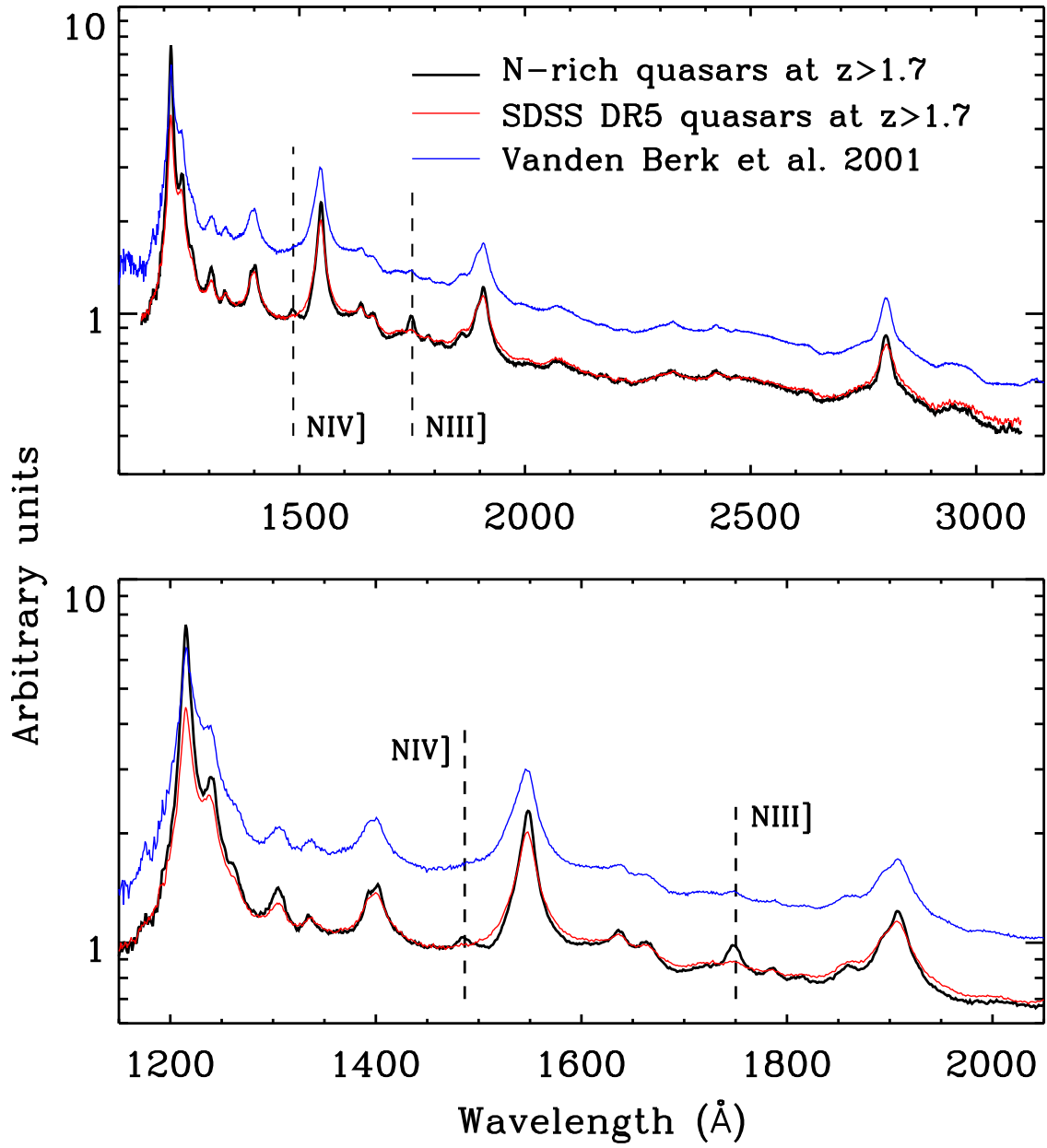}
\caption{Composite spectra for the N-rich quasars (black) and SDSS DR5
quasars (red) at $z>1.7$. The dashed lines indicate the positions of
\niv\ and \niii. For comparison, the blue lines show the composite spectrum
of \citet{van01}.}
\end{figure}

\begin{deluxetable}{ccccccccccccccc}
\tabletypesize{\scriptsize}
\tablecaption{SDSS DR5 N-rich Quasar Catalog}
\tablewidth{0pt}
\tablehead{\colhead{Name (J2000.0 Coordinates)} & \colhead{Redshift} & 
  \colhead{$i$\tablenotemark{a}} & \colhead{$\alpha$} & \colhead{} &
  \multicolumn{6}{c}{EW (\AA)} & \colhead{} & \colhead{\feii/\mgii} &
  \colhead{$M_{2500}$} & \colhead{$R$} \\
  \cline{6-11} \\
  \colhead{} & \colhead{} & \colhead{} & \colhead{} & \colhead{} &
  \colhead{\niv} & \colhead{\niii} & \colhead{\siiv} & \colhead{\civ} & 
  \colhead{\ciii} & \colhead{\mgii} & \colhead{} & \colhead{} & 
  \colhead{} & \colhead{}}
\startdata
SDSS J124032.96+674810.8& 1.701& 18.91& -0.63& & \nodata&  4.0& \nodata& 27.1& 31.2& 11.4& & 4.2& -25.27&   \nodata \\
SDSS J082247.75+071154.7& 1.703& 18.85& -0.46& & \nodata&  5.3& \nodata& 35.6& 27.1& 14.6& & 2.5& -25.30&   \nodata \\
SDSS J100643.89+395222.7& 1.705& 18.89& -0.85& & \nodata&  4.2& \nodata&101.7& 22.8& \nodata& &\nodata& -25.26&   \nodata \\
SDSS J141000.79+641010.4& 1.707& 18.46& -1.15& & \nodata&  4.1& \nodata& \nodata& 37.4& 21.8& & 3.1& -25.66&   \nodata \\
SDSS J105645.42+414016.2& 1.708& 18.71& -0.32& & \nodata&  3.0& \nodata& 26.6& 20.5&  7.8& & 4.4& -25.48&   \nodata \\
SDSS J105743.11+532231.4& 1.708& 18.52& -0.57& & \nodata&  3.4& \nodata& 41.0& 21.2& 15.9& & 2.4& -25.70&   \nodata \\
SDSS J075230.44+272619.8& 1.708& 18.25& -0.62& & \nodata&  3.3& \nodata& 67.2& 41.4& 24.4& & 1.8& -25.87&   \nodata \\
SDSS J151557.86+383604.3& 1.710& 18.24& -0.94& & \nodata&  3.6& \nodata& \nodata& 45.2& 16.9& & 2.9& -25.96&   \nodata \\
SDSS J115419.40+145555.3& 1.711& 17.98& -0.87& & \nodata&  3.6& \nodata& \nodata& 15.9& \nodata& &\nodata& -26.30&   \nodata \\
SDSS J162923.06+462311.2& 1.718& 18.95& -0.39& & \nodata&  3.2& \nodata& 43.6& 28.8& 13.8& & 2.9& -25.28&   \nodata \\
SDSS J145349.70+482740.3& 1.720& 18.81& -1.07& & \nodata&  5.1& \nodata& 19.9& 20.6& \nodata& &\nodata& -25.52&  232.6 \\
SDSS J074220.27+343213.3& 1.724& 19.77& -0.92& & \nodata&  4.8& \nodata& \nodata& \nodata& \nodata& &\nodata& -24.41&   71.4 \\
SDSS J161118.02+393350.9& 1.724& 17.82& -0.57& & \nodata&  3.9& \nodata& 30.9& 16.6& \nodata& &\nodata& -26.38&    4.3 \\
SDSS J081710.54+235223.9& 1.732& 18.61& -0.40& &  4.9&  6.7& \nodata&  8.1& 13.7& \nodata& &\nodata& -25.58& 1901.2 \\
SDSS J145958.75-013742.4& 1.738& 18.84& -1.44& & \nodata&  6.9& \nodata& \nodata& 45.1& \nodata& &\nodata& -25.35&   \nodata \\
SDSS J084057.02+054733.4& 1.742& 18.15& -1.50& & \nodata&  4.3& \nodata& 24.4& 24.3& \nodata& &\nodata& -26.25&   \nodata \\
SDSS J123354.39+150203.7& 1.744& 18.13& -1.29& & \nodata&  5.1& \nodata& 34.3& 22.8& 15.2& & 2.5& -26.18&   \nodata \\
SDSS J084117.24+083753.4& 1.746& 18.11& -1.04& & \nodata&  5.6& \nodata& 39.8& 26.5& 15.6& & 2.7& -26.17&   \nodata \\
SDSS J111559.52+633813.6& 1.746& 18.04& -1.12& & \nodata&  3.4& \nodata& 46.5& 25.3& \nodata& &\nodata& -26.15&   22.7 \\
SDSS J231608.20+132334.5& 1.748& 18.42& -1.23& & \nodata&  9.1& \nodata& 52.6& 27.5& \nodata& &\nodata& -25.84&   \nodata \\
\enddata
\tablenotetext{a}{Corrected for Galactic extinction.}
\tablecomments{The full Table 1 
appears in the electronic edition of the {\it Astrophysical Journal}.}
\end{deluxetable}

\begin{deluxetable}{lcc}
\tablecaption{Rest-frame Emission Line Properties}
\tablewidth{0pt}
\tablehead{\colhead{Parameter} & \colhead{N-rich quasars} &
  \colhead{SDSS quasars}}
\startdata
$\alpha$                   & --0.68$\pm$0.40   & --0.64$\pm$0.57  \\
\siiv\ EW (\AA)            & 8.1$\pm$3.3       & 8.1$\pm$3.4 \\
\siiv\ FWHM (\AA)          & 22.9$\pm$5.4      & 26.3$\pm$8.3 \\
\civ\ EW (\AA)             & 36.9$\pm$16.6     & 36.1$\pm$19.7 \\
\civ\ FWHM (\AA)           & 18.9$\pm$8.7      & 27.9$\pm$11.0 \\
\ciii\ EW (\AA)            & 21.8$\pm$9.4      & 23.8$\pm$9.0 \\
\ciii\ FWHM (\AA)          & 37.2$\pm$14.9     & 50.5$\pm$21.0 \\
\mgii\ EW (\AA)            & 13.7$\pm$5.0      & 11.1$\pm$4.6 \\
\mgii\ FWHM (\AA)          & 41.6$\pm$7.8      & 43.1$\pm$12.0 \\
\feii/\mgii\               & 3.6$\pm$1.3       & 3.5$\pm$1.4 \\

                           &                   &  \\
\tablenotemark{a}\lya\ EW (\AA)             & 71.0              & 50.0 \\
\tablenotemark{a}\lya\ FWHM (\AA)           & 10.6              & 15.8 \\
\tablenotemark{a}\nv\ EW (\AA)              & 24.6              & 17.8 \\
\tablenotemark{a}\nv\ FWHM (\AA)            & 18.0              & 19.0 \\
\tablenotemark{a}\oi\ EW (\AA)              & 3.7               & 1.8  \\
\tablenotemark{a}\oi\ FWHM (\AA)            & 14.8              & 14.1 \\
\tablenotemark{a}\civ\ EW (\AA)             & 38.4              & 39.6 \\
\tablenotemark{a}\civ\ FWHM (\AA)           & 18.6              & 24.0 \\
\tablenotemark{a}\ciii\ EW (\AA)            & 23.4              & 24.9 \\
\tablenotemark{a}\ciii\ FWHM (\AA)          & 35.1              & 44.9 \\
\enddata
\tablenotetext{a}{Derived from the composite spectra in $\S\,3.2$.}
\tablecomments{The errors are $1\sigma$ rms of the distributions.}
\end{deluxetable}

\end{document}